\documentclass[fleqn,usenatbib]{mnras}

\usepackage{newtxtext,newtxmath}
\usepackage{dblfloatfix}
\usepackage{float}

\newcommand{\cmt}{\textcolor{black}}
\usepackage{orcidlink}

\usepackage[T1]{fontenc}
\newcommand{\rxj}{RX-J0440.9+4431 }
\newcommand{\herx}{Hercules X-1 }

\DeclareRobustCommand{\VAN}[3]{#2}
\let\VANthebibliography\thebibliography
\def\thebibliography{\DeclareRobustCommand{\VAN}[3]{##3}\VANthebibliography}

\usepackage{graphicx}	
\usepackage{amsmath}	

\title[X-ray polarimetry timing]{Implementing and Verifying a Fourier Domain Approach to Fast Stochastic X-ray Polarimetry Timing}

\author[Melissa D. Ewing et al.]{
Melissa D. Ewing,$^{1}$\thanks{E-mail: m.ewing2@newcastle.ac.uk}\orcidlink{0000-0001-9349-8271}
Adam Ingram$^{1}$\orcidlink{0000-0002-5311-9078}
John Rankin$^2$ \orcidlink{0000-0002-9774-0560}
Fabio Muleri$^3$\orcidlink{0000-0003-3331-3794}
\\
$^{1}$School of Mathematics, Statistics, and Physics, Newcastle University, Newcastle upon Tyne, NE1 7RU, UK\\
$^{2}$INAF, Osservatorio Astronomico di Brera, Via Bianchi 46, I-23807 Merate (LC), Italy\\
$^{3}$INAF, Istituto di Astrofisica Spaziale e Fisica Cosmica, Via U. La Malfa 153, I-90146 Palermo, Italy\\
}

\date{Accepted XXX. Received YYY; in original form ZZZ}

\pubyear{\the\year{}}

\begin{document}
\label{firstpage}
\pagerange{\pageref{firstpage}--\pageref{lastpage}}
\maketitle

\begin{abstract}
The launch of the Imaging X-ray Polarimetry Explorer (IXPE), the first space-based polarimeter since 1978, offers a two order of magnitude improvement to the measurement of X-ray polarisation than its predecessor OSO-8, offering unprecedented precision for the measurement of polarisation degree and polarisation angle of X-ray sources. This advancement lends itself to the birth of a number of contemporary techniques to study Galactic compact objects, including X-ray polarimetry-timing, the study of how polarisation properties evolve over short timescales. 
However, the statistical nature of polarisation measurements poses a challenge for studies on arbitrarily short timescales, as a large number of photons are required to achieve statistically significant measurements of polarisation degree and angle for time-resolved analyses. 
Furthermore, if the polarisation variability is stochastic, then phase-folding techniques introduce systematic errors in the phase assignment of photons. Ingram and Maccarone presented a model independent Fourier-based technique that circumvents these issues.
It can be used on arbitrarily short timescales for any kind of variability, 
whether aperiodic, quasi-periodic or purely periodic. Here we implement this method on real IXPE data. We address several instrumental effects and test the technique on X-ray pulsars, \rxj and \herx. We verify that our technique recovers the polarisation variability signal that we already know to be there from typical phase-folding techniques. It will now be possible to study fast stochastic polarisation variability of X-ray sources, with applications including quasi-periodic oscillations, mass accretion rate fluctuations, and reverberation mapping.

\end{abstract}




\section{Introduction}
X-ray binaries (XRBs) are extremely X-ray bright systems powered by the accretion of matter from a companion star onto a compact object, commonly a black hole (BH) or a neutron star (NS). Transient XRBs typically remain in quiescence characterised by a low accretion rate and flux with the exception of outbursts, which transition through hard, intermediate and soft states dependent on the dominating accretion region \citep{Fender2004,Done2007,Belloni2010}. During an outburst, a source begins in the hard state, where a cloud of hot free electrons within the inner accretion flow, known as the \textit{corona} \citep{Sunyaev1985}, dominates over a geometrically thin thermalised accretion disc \citep{shakura1973black,novikov1973black} through inverse Comptonisation of disc seed photons, producing a power-law spectrum with a high-energy cut-off \citep{1976ApJ...204..187S}. Through an intermediate state which prominently displays both spectral components, the source transitions to the soft state, where now the spectrum is dominated by the thermalised disc. Finally, again through an intermediate state, the source moves back to the hard state before retuning to quiescence. Some XRBs are instead persistent, meaning that their flux does not drop to a quiescent level, but state transitions still occur.

The geometry of the corona is still under speculation as XRBs cannot be spatially resolved \citep{Poutanen2018}. Leading theories such as the `sandwich' model \citep{Galeev1979}, whereby the corona is held above and below the disc by magnetic fields; the `jet-base' model \citep{Markoff2005} in which the corona is vertically extended at the base of an outflowing jet, the `lamppost' model \citep{martocchia2002origin} where a compact corona lies above the BH on its spin axis, and the `truncated disc' model \citep{eardley1975cygnus,esin1997advection} whereby the disc evaporates beyond a truncation radius into a large scale-height corona, all reproduce the observed spectrum, and so more information is required to break the degeneracies between these models. 

XRBs are variable over many different timescales \citep{vanderKlis2006}, hence timing analyses utilising Fourier techniques can be used to break these degeneracies. Quasi-periodic oscillations (QPOs), observed as broad peaks in the power spectral density of the flux are often observed. The most common are Type-C QPOs, the origin of which has been suggested to derive from Lense-Thirring (LT) precession of the corona \citep{Ingram_2019}. 
The combination of disciplines into spectral-timing facilitates further studies. For example, it is possible to measure the time lag between flux variations in different energy bands as a function of Fourier frequency \citep{vanderKlis1987}. At low frequencies (long variability timescales), hard photons lag soft photons \citep{Miyamoto1988}, which is thought to be due to the inward propagation of mass accretion rate fluctuations \citep{Lyubarskii1997, Ingram2013}. At high frequencies, soft photons lag hard photons (e.g. \citealt{Kara2019}), which is thought to be due to the light crossing delay between X-rays observed directly from the corona and those that first reflect off the disc. These soft lags therefore enable \textit{reverberation mapping} \citep{Uttley2014}.
\textit{Polametric} studies, as made accessible by the launch of the Imaging X-ray Polarimetry Explorer (IXPE; \citealt{Weisskopf2016}) in 2021, introduce the measurement of polarisation degree (PD) and polarisation angle (PA) of XRBs, both of which are influenced by the inner accretion flow geometry.
In several sources \citep[e.g.][]{Krawczynski_2022,ingram2024trackingxraypolarizationblack} it has been found that the PA aligns with the polar radio jets, implying the corona is horizontally extended in the plane of the disc. This favours models such as the truncated disc model over those vertically extended such as the lamppost model.
Our newfound ability to measure X-ray polarisation suggests the possibility of fast stochastic X-ray polarimetry timing \citep{Ingram_2023} - i.e. tracking how the PD and PA vary on short timescales.
This will open many new avenues into the study of XRBs and the geometry of the corona. Perhaps the most obvious example is for
QPOs, since a distinguishing prediction of the LT precession model is that both the PD and PA should be modulated over the QPO period as the orientation of the corona varies over each precession cycle \citep{Fragile2025,Ingram2015a}.
It may also be possible to track accretion rate fluctuations propagating from the corona to the jet, due to synchrotron emission from the jet having a much larger PD than the Comptonised emission from the corona \citep{https://doi.org/10.48550/arxiv.2206.11671}.
Polarisation information could also enhance reverberation mapping because the direct and reflected components have different polarisation properties \citep{https://doi.org/10.48550/arxiv.2206.11671}.

However, it is technically challenging to measure changes in PD and PA on short timescales. This is because many photons are needed to detect polarisation with satisfactory statistical significance \citep[e.g.][]{Muleri_2022}, meaning that we cannot simply make a time series of PD and PA with arbitrarily short time bins. IXPE observations of a typical XRB with $\rm PD\lesssim 10\%$ do not reach the count rate required to resolve these polarisation properties for exposures $\lesssim$ 15 minutes and so the sub-second variability timescales of interest are inaccessible \citep{Ingram2017}. Periodic variability such as pulsations from X-ray pulsars can be detected with phase-folding techniques \citep[e.g.][]{Doroshenko_2022,Doroshenko_2023}. However, the stochastic nature of aperiodic variability prohibits traditional phase-folding techniques used to detect periodic variability. Although techniques do exist to phase-fold QPOs \citep{huang2014hilbert, Henric2016,Zhao_2024}, they rely on assumptions that systematically affect the results, and cannot be applied to all broadband aperiodic variability.

\cite{10.1093/mnras/stx1881} presented a Fourier-based method to
detect
fast stochastic polarisation variability.
It can be used on arbitrarily short timescales over all kinds of variability whether aperiodic, quasi-periodic or purely periodic. Here we implement the technique on real IXPE data for the first time. We address several instrumental effects and test the technique on X-ray pulsars. We verify that our technique recovers the polarisation variability signal that we already know to be there from traditional phase-folding techniques.

We choose two sources with contrasting spin periods,  \rxj and \herx \citep{Doroshenko_2023,Doroshenko_2022}. \rxj is a Be XRB discovered in 1996 \citep{motch1996} at a distance of $\approx2.4 \pm 0.1$kpc with a spin period of $T_{\rm spin}\simeq24$s. \herx was one of the first X-ray pulsars to ever be discovered \citep{tananbaum1972}, with a distance of $\approx7$ kpc and a spin period $T_{\rm spin}\simeq 1.24$s.\\
\\
In Section \ref{sec:Observations}, we detail the data reduction procedure. In Section \ref{sec:Methods}, we outline the \cite{10.1093/mnras/stx1881} Fourier-based method and describe the instrumental effects and how we circumvent them. In Section \ref{sec:Results} we discuss the results of applying the method to X-ray pulsars \rxj and Hercules X-1. In Section \ref{sec: sims}, we present simulations characterising instrumental effects whereby we `re-inject' the observed polarisation variability. In Section \ref{sec:Conclusions}, we present our conclusions.
\section{Observations}
\label{sec:Observations}
\subsection{Data reduction and selection}
IXPE is a joint NASA/Italian Space agency mission launched  on December $9^{\rm th}$ 2021 from the Kennedy Space Center. It consists of three identical  gas pixel detector units (DUs) which record the spatial, energy, timing, and polametric information from each event within a $2{-}8$keV energy band. Specifications and observatory details can be found in \cite{2022JATIS...8b6002W}. IXPE has observed both \rxj and \herx several times, so we select the longest observation in each case. The \rxj observation (OBSID:02250501) took place between $23^{\rm rd}$ Feb - $3^{\rm rd}$ March 2023 with $\sim$8.3M source counts. The \herx observation (OBSID:02004001) took place between Feburary $3^{\rm rd}$-$8^{\rm th}$ 2023 with $\sim$1.5M source counts. We selected source photons within a circular source region of radius 60" and did not subtract a background count rate due to the high source count rates in each observation as advised in \cite{Di_Marco_2023}.\\
\\
We combined the good time intervals (GTIs) from the three DUs into a master GTI list by specifying that all three DUs must be active during a GTI. We then filtered all events on our new GTI list.
We corrected the photon arrival times of each observation to the Solar System barycentre using the \texttt{barycorr} task \citep{2014ascl.soft08004N}, applying the latest planetery ephemredies (JPLEPH.430). We additionally corrected the \herx photon arrival times for binary orbital motion using the \texttt{binarycorr} task \citep{binarycorr}, using ephemredies found in \cite{refId0}. This correction was not required for RX-J0440.9+4431, as the effects are negligible due to its long orbital period.

\subsection{Phase-folding}
\begin{figure}
    \centering
    \includegraphics[width=1\linewidth]{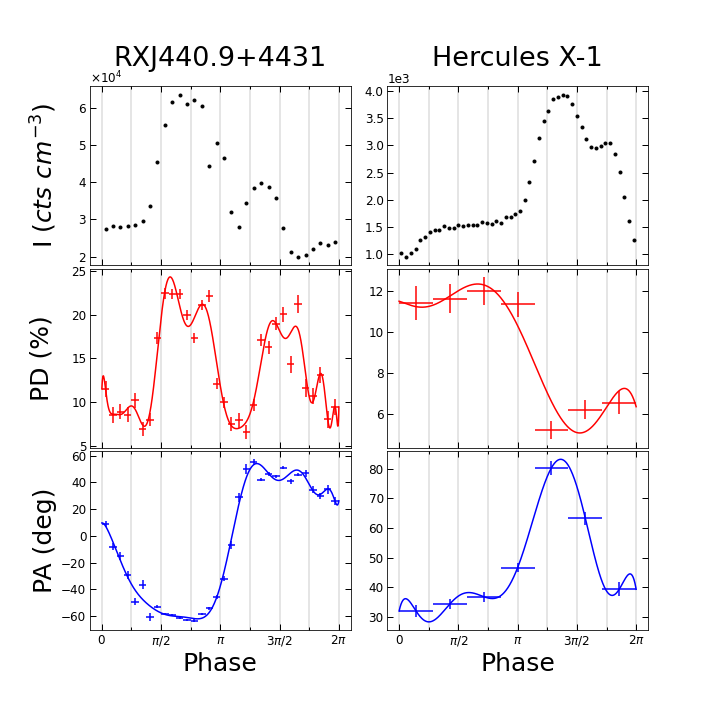}
    \caption{Pulse phase dependence of the source flux, PD and PA of RX-J0440.9+443 and Hercules X-1. The pulse profiles are in agreement with \protect\cite{Doroshenko_2022,Doroshenko_2023}. The solid lines are the polynomials used to map PD and PA to individual events for simulations (see Section~\protect\ref{sec: sims}).}

    \label{fig:phase fold}
\end{figure}

Due to the periodic nature of X-ray pulsar pulsations, we can use phase-folding techniques to recover variability in flux, PD and PA. We took the pulsar ephemredies of \rxj directly from \cite{Doroshenko_2023} which were calculated using the phase-connection technique of \cite{1981ApJ...247.1003D}.
For Hercules X-1, we started from the ephemredies quoted in \cite{Doroshenko_2022}, then refined the spin frequency to a value just outside of the $1\sigma$ limit stated in \cite{Doroshenko_2022}. We determined out revised spin frequency by maximising the folded pulse profile amplitude. Our adopted ephemredies can be found in Table \ref{table:1}.\\
\\
Events were taken between an interval of $2{-}8\,$ keV and $2{-}7\,$ keV for \rxj and \herx respectively, the latter with a smaller range to match that used in \cite{Doroshenko_2022}. 
We used the ephemredies to fold Stokes I, Q and U to produce the phase resolved PD and PA (Fig \ref{fig:phase fold}). We do not employ track weighting and calibrate with unweighted rmf and arf files. Our results are consistent with \cite{Doroshenko_2022} and \cite{Doroshenko_2023}, and demonstrate with high statistical significance that intrinsic, periodic variability exists in the PD and PA of both sources. With use of our Fourier-based method, we can move on to the `re-detection' of this result, thus demonstrating that we come to same conclusion without requiring pure periodicity of the signal.

\begin{table}
\caption{Ephemredies of \rxj and \herx used in phase phase folding (see Figure \ref{fig:phase fold}). $t_0$ is measured in seconds after the IXPE Time epoch \citep{ixpe_data_formats_2022}}
\begin{tabular}{|l|l|l|}
\hline
      & \rxj        & \herx            \\ \hline
$\nu \, \,({\rm s}^{-1}) $     & 0.004867   & 0.8079478799630468 \\ \hline
$\dot \nu\, \,({\rm s}^{-2})$  & $1.42 \times 10^{-11}$    & 0                  \\ \hline
$\ddot \nu\, \,({\rm s}^{-3})$ & $-8 \times 10^{-18}$      & 0                  \\ \hline
$t_0\, \,({\rm s})$  & 193938354.36799997 & 192238465.0        \\ \hline
    
\end{tabular}
\label{table:1}
\end{table}

\section{Methods}
\label{sec:Methods}
First presented in \cite{10.1093/mnras/stx1881} and \cite{https://doi.org/10.48550/arxiv.2206.11671}, the following section outlines our model-independent, Fourier-based method for detecting fast stochastic polarisation variability in X-ray sources. In Section \ref{sec:sum of fb method},
we first discuss the intrinsic variability imprinted on the response of IXPE as a consequence of flux, PD and PA variability, and then outline the method in which we quantify this variability utilising Fourier-based techniques. In Sections \ref{sec:deadtime}-\ref{sec: mod factor variability}, we then discuss individual challenges associated with instrumental effects, how they bias the results and the procedures we use to circumvent or account for them. Effects include those common to other X-ray missions (dead time, Section \ref{sec:deadtime}), and those unique to X-ray polarimeters (Sections \ref{sec:spur pol}-\ref{sec: mod factor variability}).

\subsection{Summary of the Fourier-based method}
\label{sec:sum of fb method}
\subsubsection{Variability in the modulation function}
\label{sec:modfunc var}
\begin{figure*}

    \centering
    \begin{minipage}{0.49\textwidth}
        \centering
        \vspace{5mm}
        \includegraphics[width=1\linewidth]{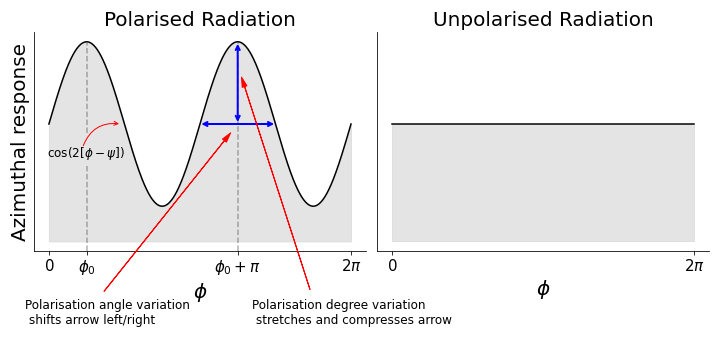}
        \vspace{5mm}
        \small (a)
    \end{minipage}%
    \hfill
    \begin{minipage}{0.49\textwidth}
        \centering
        \includegraphics[width=1\linewidth,height=0.5\linewidth]{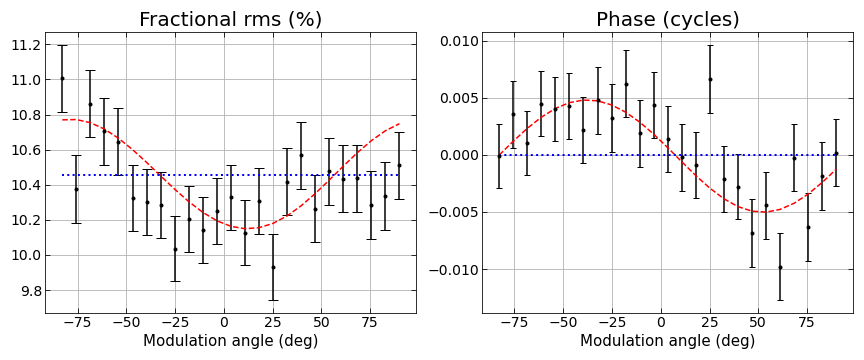}
        \small (b)
    \end{minipage}
        \caption{(a) The left plot demonstrates the modulation function in the presence of polarisation. The amplitude of the $\cos^2$ function increases with increasing polarisation degree and the position of the peak on the modulation angle ($\phi$) axis corresponds to the polarisation angle. The right plot demonstrates an idealised modulation function in the absence of polarisation. There is no dependence on counts as a function of modulation angle. (b) A simulation of the expected fractional rms and phase as a function of modulation angle \citep{Ingram2015}. The red dotted line is the best fitting sinusoidal model and the blue dashed line is the null hypothesis. Polarisation variability is detected when we statistically prefer the sinusoidal model over the flat response.}
        \label{fig:modulation function}
\end{figure*}

Our Fourier based method of X-ray polarimetry timing measures the variability of the intrinsic detector response of IXPE, known as the \textit{modulation function} in response to polarisation variability. The modulation function is proportional to a histogram of source counts  as a function of \textit{modulation angle}, $\phi$, which defines 
the direction of the photo-electron track produced by an incoming photon as it travels through the sensitive volume of the gas pixel DU. The modulation angle is cyclical over a $-90^\circ$ to $90^\circ$ interval and is correlated with the electric field direction of the absorbed photon \citep{Muleri_2022}, encoding its polarisation properties. The modulation function is given by:
\begin{equation}
    f(\phi|\psi,p,\mu) = \frac{1}{\pi}\{1 + \mu p \cos[2(\psi-\phi)]\}
    \label{eq: modulation function}
\end{equation}
where $\mu$ is the modulation factor describing the quality of the photo-electron track, $\psi$ is the PA and $p$ is the PD.\footnote{Note that $\phi$ can alternatively be defined on a $-180\degr$ to $180\degr$ interval (even though it is always \textit{cyclical} on a $-90\degr$ to $90\degr$ interval), in which case the modulation function includes an extra factor of 2 in the denominator.} \\
\\
The modulation function is dependent on flux, PD and PA, and so if these parameters are variable, then this will impart distinct changes onto its shape. As seen in Fig \ref{fig:modulation function}, PD variability will alter its amplitude, and PA variability will shift its peak. Measuring these changes over the full range of modulation angles provides an indirect method of identifying polarisation variability of any form. Note that a change in the source spectrum may also change the amplitude of the modulation function. This is investigated in Section \ref{sec: sims}.\\
\\
To isolate this variability, we employ similar techniques used in spectral timing, and create `subject' band light curves selected over a given modulation angle bin such that the count rate in the $j^{\rm th}$ bin is 

\begin{equation}
    s(\phi_j,t) = s(t)f(\phi_j|\psi(t),p(t),\mu)\Delta \phi_j,
    \label{eq:dc lc}
\end{equation}
where s(t) is the total polarimeter count rate, $f$ is the modulation function, and $\Delta\phi$ is the width of the modulation angle bin. In spectral timing, these light curves are instead binned over energy \citep{2014A&ARv..22...72U}. The properties of these light curves are therefore also dependent on changes in the flux, PD and PA. If only the PD is variable, the subject-band light curves will exhibit a different fractional variability amplitude to one another, where we observe a maximum rms at $\phi=\psi$ and a minimum rms at $\phi=\psi+90^\circ$.
If only the PA is variable then the subject-band light curves will exhibit different phases to one another, where the maximum phase lag between light curves occurs for those selected on $\phi>\psi$ and $\phi<\psi$. When both PD and PA vary together, unique shapes will be created in the modulation function such that both the fractional rms and phase vary between the subject-band light curves. In this case there will be a sinusoidal dependence on fractional rms and phase as a function of modulation angle \cite{10.1093/mnras/stx1881}. Measuring this fractional rms and phase utilises Fourier-based techniques.

\subsubsection{The Fourier-based method}
To calculate the fractional rms and phase lags between the subject-band light curves, we calculate a span of \textit{cross spectra} with respect to a `reference' band light curve, $r(t)$, over all modulation angles, given by
\begin{equation}
    C(\phi_j,\nu,\Delta) = \langle S(\phi_j,\nu)R^*(\nu) \rangle.
    \label{eq:cross spec}
\end{equation}
Here, an uppercase denotes the Fourier transform, $^*$ denotes the complex conjugate and the angle brackets denote averaging first over many realisations by splitting the light curves into segments of length $T_{\rm seg}$, then over Fourier frequency range $\nu_{\rm min}=\nu - \Delta/2$ to $\nu_{\rm max} = +\Delta/2$ (see eg. \citealt{1989ARA&A..27..517V}). We adopt squared fractional rms normalisation throughout \citep{1994A&A...283.1037B}. The fractional rms and phase are defined as:
\begin{equation}
    \text{Fractional rms} =|G|\frac{\sqrt{\Delta}}{\sqrt{\langle P_r(\nu,\Delta) \rangle}}, \hspace{5mm} \text{Phase lag}= \arg{(G)},
    \label{eq:rmsphase}
\end{equation}
where 
\begin{equation}
    \label{eq: pr}
    P_r(\nu) =\langle R(\nu)R^*(\nu)\rangle
\end{equation}
is the Poisson noise subtracted reference band power spectrum.
We employ errors on the cross spectrum as detailed in \cite{10.1093/mnras/stz2409} and propagate accordingly.






\subsection{Deadtime}
\label{sec:deadtime}

The effect of deadtime \citep{Bachetti2015} - the time it takes a detector to become responsive after recording an event -
results in the sensitivity of a detector varying in antiphase with the incident count rate. This in turn induces a $\sim\pi$ radian phase lag between modulation angle bins in each of the GPD polarimeters. Furthermore, it has the effect of altering the shape of the Poisson noise in a power spectrum, imprinting on it a sinc-function shape. IXPE has a relatively high deadtime of $\sim1.24\rm \, ms$ and therefore this effect must be handled appropriately.\\ 
\\
Following \cite{Bachetti2015}, we mitigate for deadtime
by ensuring that events between subject-band and reference-band light curves are independent. 
To do this, we select events from DU1 and DU2 to make up the subject-band, and DU3 for the reference band throughout cross spectra calculation. Our choice to sum over two DUs for the subject band and only one DU for the reference band was motivated by maximising signal to noise, since \cite{10.1093/mnras/stx1881} showed that, for a given total count rate, the highest signal to noise is achieved for equal counts in the subject and reference bands. We also estimate the reference band power spectrum $P_r(\nu)$ from the co-spectrum (the real part of the cross spectrum) between the reference band light curve extracted from DU1 and DU2 and that extracted from DU3. \\
\\
Our use of independent detectors circumvents the main issues of spurious phase lags and biased Poisson noise. It does not correct for the bias in measured Fourier amplitudes caused by deadtime \citep{Huppenkothen_2018}, but this effect is small for the count rates and frequency ranges we are considering here.




\subsection{Spurious Polarisation}
\label{sec:spur pol}
The IXPE detectors experience \textit{spurious polarisation} \citep{Rankin2022}. That is, before calibration, the IXPE DUs measure a small polarisation even from an unpolarised source. Note that \citet{Rankin2022} refer to the same effect as `spurious modulation', but here we reserve the word `modulation' to mean a modulation in time. In this Section, we first introduce the concept of spurious polarisation (Section \ref{sec:speffect}), then describe how it is corrected for in the IXPE pipeline (Section \ref{sec:spcorrect}), before describing how it influences our polarimetry-timing technique and how we correct for it (Section \ref{sec:sptim}).

\subsubsection{The effect of spurious polarisation}
\vspace{1mm}
\label{sec:speffect}
In the absence of any spurious polarisation, the modulation function (Equation \ref{eq: modulation function}) can be re-written in terms of normalised Stokes parameters $q$ and $u$
\begin{equation}
f(\phi) = \frac{1}{\pi} \bigg[ 1 + q \cos( 2\phi ) + u \sin( 2\phi ) \bigg],
\end{equation}
where
\begin{equation}
q = \mu p \cos(2\psi);~~~~~ u = \mu p \sin(2\psi).
\end{equation}
The simplest way to measure polarisation is to define per-event Stokes parameters
\begin{equation}
\label{eq: 7}
q(k) = 2 \cos(2\phi_k);~~~~~~~~ u(k) = 2 \sin(2\phi_k),
\end{equation}
where $\phi_k$ is the measured modulation angle of the $k^{\rm th}$ event. The normalised Stokes parameters for a large number of events $N$ are
\begin{equation}
q = \frac{1}{N} \sum_{k=1}^{N} q(k);~~~~~~u = \frac{1}{N} \sum_{k=1}^{N} u(k),
\label{eqn:qandu}
\end{equation}
and the PD and PA are \footnote{where the phase ambiguity in the $\arctan$ function must be properly accounted for (e.g. using the \texttt{atan2(y,x)} function in most programming languages).}
\begin{equation}
    p = \frac{1}{\mu} \sqrt{ q^2 + u^2 };~~~~~~ \psi = \frac{1}{2}\arctan\left( \frac{u}{q} \right).
    \label{eqn:pandpsi}
\end{equation}

In the presence of spurious polarisation, the modulation function becomes \citep{Rankin2022}
\begin{equation}
f(\phi) = \frac{1}{\pi} \bigg[ 1 + q_{\rm tot} \cos( 2\phi ) + u_{\rm tot} \sin( 2\phi ) \bigg],
\label{eqn:fullf}
\end{equation}
where
\begin{equation}
q_{\rm tot} = q + q_{\rm sp};~~~~~u_{\rm tot} = u + u_{\rm sp},
\label{eq:qutot}
\end{equation}
and $q_{\rm sp}$ and $u_{\rm sp}$ are the normalised Stokes parameters associated with the spurious polarisation. The modulation angles IXPE measures for each event, $\phi_k$, are therefore influenced by spurious polarisation. The per-event Stokes parameters thus become
\begin{equation}
q_{\rm tot}(k) = 2 \cos(2\phi_k);~~~~~~~~ u_{\rm tot}(k) = 2 \sin(2\phi_k),
\end{equation}
and the normalised Stokes parameters become
\begin{equation}
q_{\rm tot} = \frac{1}{N} \sum_{k=1}^{N} q_{\rm tot}(k);~~~~~~u_{\rm tot} = \frac{1}{N} \sum_{k=1}^{N} u_{\rm tot}(k).
\end{equation}
If we were to ignore spurious polarisation (i.e. by assuming $q=q_{\rm tot}$ and $u=u_{\rm tot}$), Equation \ref{eqn:pandpsi} would return a \textit{biased} measurement of the PD and PA. For example, if the source were in reality unpolarised, we would erroneously measure a PD of $p=\sqrt{q_{\rm sp}^2+u_{\rm sp}^2}/\mu$.

\subsubsection{Correcting for spurious polarisation}
\vspace{1mm}
\label{sec:spcorrect}

There are two ways to correct for spurious polarisation. One is to subtract normalised Stokes parameters
\begin{equation}
    q = q_{\rm tot} - q_{\rm sp};~~~~~~~~ u = u_{\rm tot} - u_{\rm sp},
\end{equation}
where
\begin{equation}
q_{\rm sp} = \frac{1}{N} \sum_{k=1}^{N} q_{\rm sp}(k);~~~~~~u_{\rm sp} = \frac{1}{N} \sum_{k=1}^{N} u_{\rm sp}(k).
\label{eqn:qsp}
\end{equation}
\cite{Rankin2022} refer to this method as `global decoupling'\footnote{Note that global decoupling requires the spurious Stokes parameters to be extracted from the same region of the detector as the true events. For each real event $k$, the spurious polarisation Stokes parameters are determined from the DU, chip position and energy channel associated with the real event.}.
The other method is to instead subtract per-event Stokes parameters
\begin{equation}
    q(k) = q_{\rm tot}(k) - q_{\rm sp}(k);~~~~~~~~u(k) = u_{\rm tot}(k) - u_{\rm sp}(k).
    \label{eqn:qucorr}
\end{equation}

In this case, we can simply calculate the PD and PA using Equations \ref{eqn:qandu} and \ref{eqn:pandpsi}. This second method is the one employed in the IXPE pipeline (see Appendix \ref{sec:ixpeSP} for further details). It is preferable to global decoupling because it enables spurious polarisation to be eliminated already in the event files, such that a typical end-user never needs to even be aware of spurious polarisation \citep{Rankin2022}. Thus, the Level 2 IXPE event files list the corrected per-event Stokes parameters $q(k)$ and $u(k)$ from Equation \ref{eqn:qucorr}, and not the per-event modulation angles $\phi_k$ \citep{Muleri_2022}.

\subsubsection{Relevance to polarimetry-timing}
\vspace{1mm}
\label{sec:sptim}

For our polarimetry-timing technique we need to recover the modulation angles, which are not listed in the Level 2 event files. This is not trivial because we need the modulation angles in the \textit{sky} frame, whereas the modulation angles listed in the Level 1 event files are instead in the \textit{detector} frame. The IXPE pipeline converts from detector to sky frame (by rotating the detector frame modulation angle by a time-dependent \textit{roll angle} $\Delta_k$) \textit{and} corrects for spurious polarisation in the same step from Level 1 event files to Level 2 event files. We therefore need to define our own intermediate step in the pipeline that we refer to as `Level 1.5'. Our code produces Level 1.5 event files that list for each event $\phi_k$, $q_{\rm sp}(k)$ and $u_{\rm sp}(k)$ as well as $q(k)$ and $u(k)$. We describe our code in Appendix \ref{sec:Deltamethod}.\\
\\
We can then create $\phi-$selected light curves as required for our polarimetry-timing technique. To do this we extract $J$ light curves, each selected for modulation angles in the range $\phi_j-\Delta\phi/2$ to $\phi_j+\Delta\phi/2$, where $\Delta\phi = \pi/J$. Each light curve has $I$ time steps of duration $dt$, such that the number of counts in the $i^{\rm th}$ time bin of the $j^{\rm th}$ light curve is $N(t_i,\phi_j)$. The total number of counts is therefore $N = \sum_{j=1}^J \sum_{i=1}^I N(t_i,\phi_j)$. However, we now need to correct for spurious polarisation. The effects of constant spurious polarisation can be accounted for using the global decoupling method described in Section \ref{sec:spcorrect}. We apply such a correction following
\begin{equation}
N_{\rm corr}(t_i,\phi_j) = N(t_i,\phi_j) - \frac{N}{I J} \bigg[ q_{\rm sp} \cos(2\phi_j) + u_{\rm sp} \sin(2\phi_j) \bigg].
\label{eq:ncorr}
\end{equation}
The above is an exact correction \textit{if} the spurious polarisation is constant (the derivation is presented in Appendix \ref{sec:ncorr}).

However in reality the contribution of spurious polarisation can vary in time. This is because the IXPE spurious polarisation depends on photon energy and position on the chip. These dependencies are all constant in time, but variability in spurious polarisation can be caused by the source moving across the chip with time and by the spectrum of the source changing with time. IXPE is dithered over a small circular region (radius=1.5mm) to facilitate calibration, which induces variability in the spurious polarisation Stokes parameters on the dithering periods (107, 127 and 900 s) as source photons are processed on different areas of the chip at different times. Moreover, since the amount of spurious polarisation decreases with photon energy, epochs with a softer spectrum will be associated with more spurious polarisation than epochs with a harder spectrum. It is common for the spectral shape to vary systematically with QPO phase \citep[e.g.][]{Ingram2016} and with pulse phase \citep[e.g.][]{Miura_2024}, which will drive a small amount of spurious polarisation variability on the QPO/pulse phase. The dominant source of spurious polarisation variability is therefore timescale dependent: on the dithering period the dominant source is chip position, and on the QPO/pulse period the dominant source is energy dependence. To quantify this variability, we develop a simulation pipeline, which is described in Section \ref{sec: sims}.

\subsection{Modulation factor variability}
\label{sec: mod factor variability}
The modulation factor, $\mu$, as seen in Equation \ref{eq: modulation function}, is a measure of how effectively we can measure polarisation properties for each recorded event. It is a function of photon energy since higher energy photon tracks are more easily recovered resulting in a greater $\mu$ \citep{Muleri_2022}. Since the source spectrum may vary over the variability timescale being considered (e.g the QPO or pulse period), variability in the modulation factor will induce rms variability, the same effect to that as a variable PD.
We test the affect of this through simulations (see Section \ref{sec: sims}).

\section{Results}
\label{sec:Results}
In this section, we present the results of applying our Fourier-based method to sources \rxj and \herx with null hypothesis significance testing. We created subject and reference band light curves with a time resolution of $dt=1/64 \, s$, and produced a total of 20 subject band light curves.
We calculated the cross-spectrum between subject and reference band light curves in squared fractional rms normalisation, and averaged over segment lengths of 1024s and 16s for \rxj and \herx respectively. We then averaged over the fundamental pulse frequency of each source corresponding to the Fourier frequency ranges $(0.00390625{-}0.005859375)$Hz and $(0.7{-}0.9)$Hz for \rxj and \herx respectively. We calculated the fractional rms and phase from Equation \ref{eq:rmsphase} and \cite{Ingram2019} error bars were employed.\\
\begin{figure*}
    \centering
    \begin{minipage}{0.49\textwidth}
        \centering
        \includegraphics[width=1\linewidth]{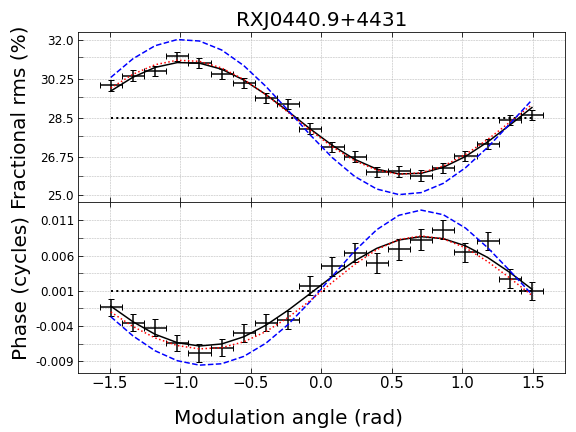}
        
    \end{minipage}%
    \hfill
    \begin{minipage}{0.49\textwidth}
        \centering
        \includegraphics[width=1\linewidth]{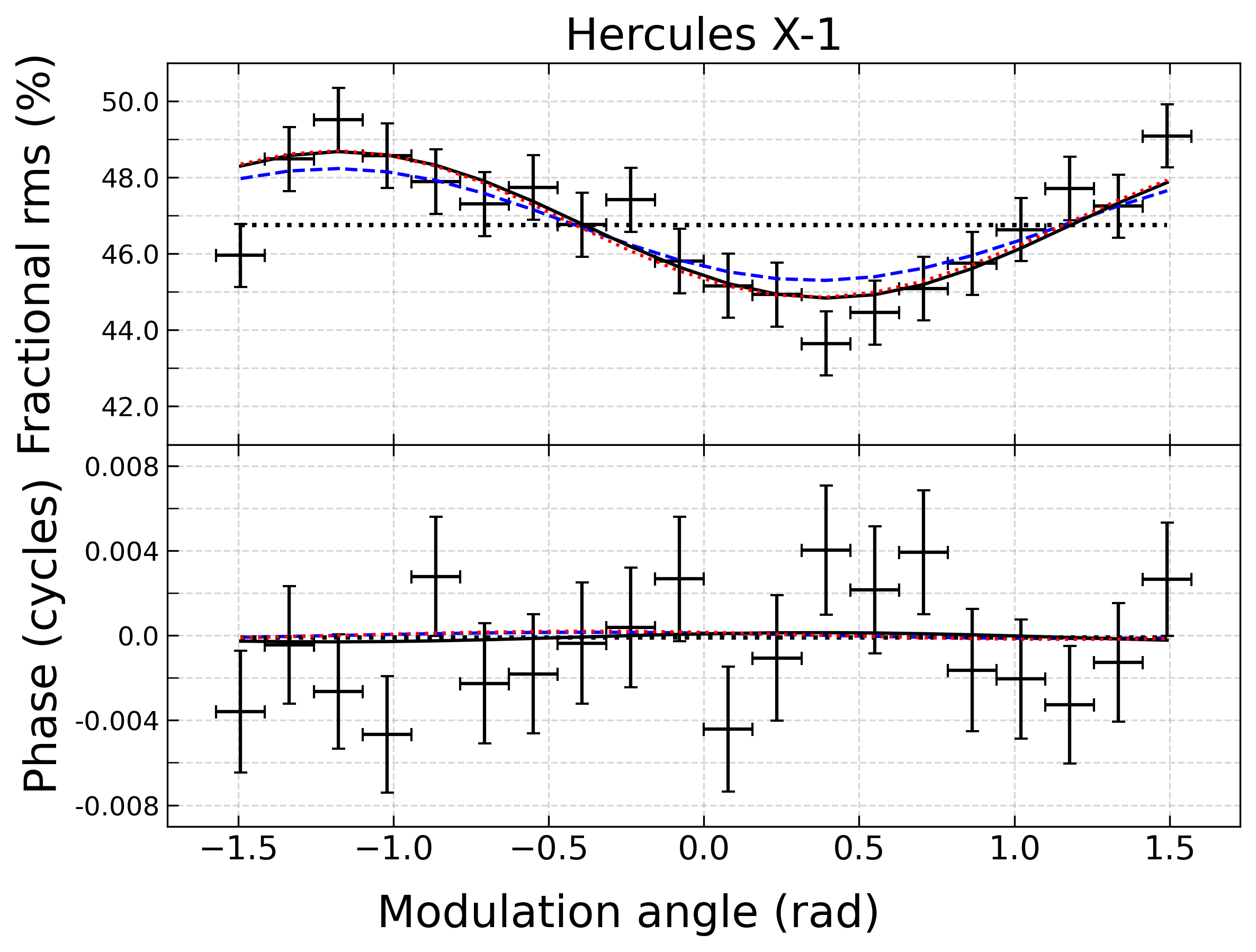}
        
    \end{minipage}
    
    \caption{Observed fractional rms (upper plot) and phase (lower plot) as a function of modulation angle of \rxj (left) and \herx (right). The black solid lines are the best fitting sinusoidal models and the black dotted lines are the best fitting null hypothesis model.
    \cmt{The remaining two lines represent the rms and phase calculated analytically from the phase folded light curves. For the blue dashed lines, we use values of the modulation factor calculated from the raw data. For the red dotted lines, we instead choose values for the modulation factor that optimise agreement with the observed rms and phase.}}
    \label{fig:real results}
\end{figure*}
\\
Figure \ref{fig:real results} shows the fractional rms (top) and phase lag (bottom) as a function of modulation angle over the fundamental pulse frequency of \rxj (left) and \herx (right) respectively. 
\subsection{Null-hypothesis testing}
If the PD and PA were constant with pulse phase, then both the fractional rms and phase would be independent of modulation angle. Instead we see sinusoidal dependencies. For \rxj, we see a particularly strong effect, but it is also clear in the rms of Hercules X-1. 
To determine a detection significance, we fit the data with a sine model
\begin{equation}
   y= A + B\sin[(2(\phi+C)]
   \label{eq: 180mod}
\end{equation}
where A, B and C are free parameters. The solid black lines show the best fitting model. The dashed black lines show the best fitting constant, i.e. the null hypothesis model. Our model gives $\chi^2$ values of 25.07
and 33.58
for 34 degrees of freedom (40 data points and 6 free parameters), for \rxj and \herx respectively. The null hypothesis gives $\chi^2$ values of 1600
and 86.36
for 38 degrees of freedom (40 data points and 2 free parameters). An F-test returns F-statistics of
534.1 and 13.87,
corresponding to a statistical significance of  $>8\sigma$
and 
$4.93\sigma$
in preference of the sinusoidal model over the null hypothesis. We therefore conclude that the Fourier based method successfully recovers the present polarisation variability.\\
\\
As expected, \rxj demonstrates a more significant positive detection than \herx due to the observation's large count number and longer spin period.

\subsection{Analytical calculation}
\label{sec:analytic}
In this section, we test the validity of our model and our methods of dealing with instrumental effects by calculating the fractional rms and phase analytically from the phase folded light curves in Figure \ref{fig:phase fold} to compare to the observed data. A strong agreement between observed and analytic data consolidates that our method works as expected and that we have corrected for instrumental effects appropriately. To do this we calculate an expected modulation function (with 20 modulation angle bins) for each of our pulse phase bins (32 and 7 for \rxj and \herx respectively) and carry out our Fourier-based analysis as usual. We start by scaling each pulse profile to match the fractional rms calculated from the observed data in the pulse frequency ranges, and then calculate the expected modulation function for each pulse phase bin by substituting the phase folded PD and PA into Equation \ref{eq: modulation function}.

We set the modulation factor in our calculation to the effective modulation factor, calculated via $\mu=\sqrt{q_{\rm uncorr}^2+u_{\rm uncorr}^2}/PD$, where $q_{\rm uncorr}$ and $u_{\rm uncorr}$ are the normalised Stokes parameters \textit{not} corrected for modulation factor (i.e. the stokes parameters for each event have not been divided by the corresponding modulation factor). This returns $\mu=0.286$ and $\mu=0.336$ for \rxj and \herx respectively. We note that, although our analytic calculation assumes $\mu$ to be constant, in reality the effective modulation factor will vary with a small amplitude (see Section \ref{sec: mod factor variability}). We consider the effect of this discrepancy in Section \ref{sec: sims}.


We generate 20 subject band light curves by substituting the calculated modulation function and phase-folded flux into Equation \ref{eq:dc lc}, and for reference band light curve we use the true phase folded light curve. We then carry out the Fourier-based method on these light curves just as we do with the raw data, utilising Equations \ref{eq:cross spec} - \ref{eq: pr}.

The blue dashed lines in Figure \ref{fig:real results} show the analytically calculated fractional rms and phase as a function of modulation angle for each source. For \rxj we find that the analytical calculation slightly over-predicts the observed fractional rms and phase variability, whereas for \herx we find agreement between the two within errors. \cmt{For the red dotted lines, we re-do the analytical calculation except now we choose values of the modulation factor $\mu$ that optimise agreement with the observational data. The values used are $\mu=0.21$ and $\mu=0.44$ for \rxj and \herx respectively. We see that the discrepancy with the data can be eliminated entirely by adjusting the assumed modulation factor. Out initial calculation of the effective modulation factor may be sub-optimal because it ignores non-linear effects. In the real data, the modulation factor effectively varies in time, and when we calculate our Fourier products we isolate a small range of timescales around the pulse period. In contrast, our analytical calculation takes Stokes parameters folded on the pulse phase (isolating variability on the pulse phase and all of its overtones) and a time-averaged modulation factor, where the averaging is over all timescales. In the following section, we test this hypothesis by running a simulation that automatically accounts for all non-linear effects.}
We performed the pulse profile scaling such that the rms amplitude of the pulse fundamental matched the rms amplitude of the unfolded light curve integrated over the frequency range $\nu_{\rm lo}$ to $\nu_{\rm hi}$. This rescaling was necessary for two reasons. First, the unfolded light curves include variability not associated with the pulse. This effect on its own would lead to the fundamental rms of the folded light curve being less than that of the unfolded light curve, which in turn would lead to the analytical prediction under-predicting the rms by a constant offset. We find that this first effect is dominant for \herx, and thus need to scale the pulse amplitude by 1.2847 in our calculation. Second, the use of discrete Fourier transforms spreads the pulse variability over a range of Fourier frequencies, in contrast to phase-folding. This effect on its own would lead to the rms of the folded light curve being greater than that estimated from the unfolded light curve. We find that this second effect is dominant for RX-J0440.9+4431, and thus we need to scale the pulse amplitude by 0.88256.
\section{Simulations}
\label{sec: sims}
In this section, we present the results of applying the Fourier-based method to simulated data replicating the observed polarisation properties of each source. This confirms that we reproduce the observed fractional rms and phase as a function of modulation angle whilst simultaneously testing the methods used to account for the instrumental effects which we cannot simply circumvent, namely spurious polarisation and modulation factor variability.
For each simulation, we use real events from the Level 1.5 event files. We keep the true arrival time, energy channel, and chip position of each event. This enables us to simulate the polarisation properties, whilst preserving all other properties of the real IXPE data. From the resulting event lists, we extract a set of simulated subject-band light curves (note that the reference band light curve is exactly the same as that of the real data) and carry out the Fourier-based method as usual, utilising Equations \ref{eq:dc lc} through \ref{eq: pr}. We repeat this process many times, each time generating new modulation angles for each event. We then average the rms and phase lags over all iterations to minimise the statistical noise associated with selecting random variables.

To generate a simulated modulation angle per event, we first calculate the pulse phase of the event from the time of arrival and pulsar ephemerides, then assign a corresponding PD and PA according to the phase folded light curves in Figure \ref{fig:phase fold}. Since the PD/PA phase folded light curves are discrete, (i.e. polarisation between pulse phase bins is not defined), we use best-fitting polynomials to map all ranges of pulse phase to the corresponding polarisation properties (solid lines in Figure \ref{fig:phase fold}). The probability of an event having a modulation angle $\phi$ with polarisation PD and PA is given by the modulation function $f(\phi| p,\psi)$ (Equation \ref{eq: modulation function}) and so our simulated modulation angle of each event is a random value drawn from the modulation function. We randomly sample from this distribution by implementing the Monte-Carlo Inverse transformation method, setting the modulation function as the probability distribution function and recovering the angle using Halley's method.

To test the effects of spurious polarisation, we carry out two simulation scenarios: one where we account for spurious polarisation and one where we ignore it. In the case where we ignore it, we took the PD and PA per event to be exactly that mapped from the phase folded light curves. In the case where we incorporate the effects of spurious polarisation we recalculate PD and PA values that include spurious polarisation. We did this by first calculating the initial stokes q and u from the mapped PD and PA
$q=\mu p\cos{(2\psi)}$ and $u=\mu p\sin{(2\psi)}$, where $\mu$ is selected for the energy channel of the real event.
We then add on spurious stokes parameters using Equation \ref{eq:qutot}. The new PD and PA are then given by 
$p_{\rm tot}=\sqrt{q_{\rm tot}^{2}+u_{\rm tot}^{2}} / \mu$ and $\psi_{\rm tot}=\frac{1}{2}\arctan(\frac{u_{\rm tot}}{q_{\rm tot}})$ respectively.

To test the effects of a variable modulation factor, we again run two simulation scenarios where in one case, we use the true modulation factor assigned per event according to its energy channel, and in the other case, we set it to the respective effective modulation factors used in the analytical calculation, so we can further compare these simulations against the analytic model.
To test the interplay between these two instrumental effects we run four sets of simulations for each source to test all combinations of spurious polarisation (considered / not considered) and modulation factor variability (constant / not constant).
For each scenario of each source, we simulate 100 observations and average over the resulting fractional rms and phase.
\begin{figure*}
    \centering
    \begin{minipage}{0.49\textwidth}
    
        \centering
        \includegraphics[width=1\linewidth]{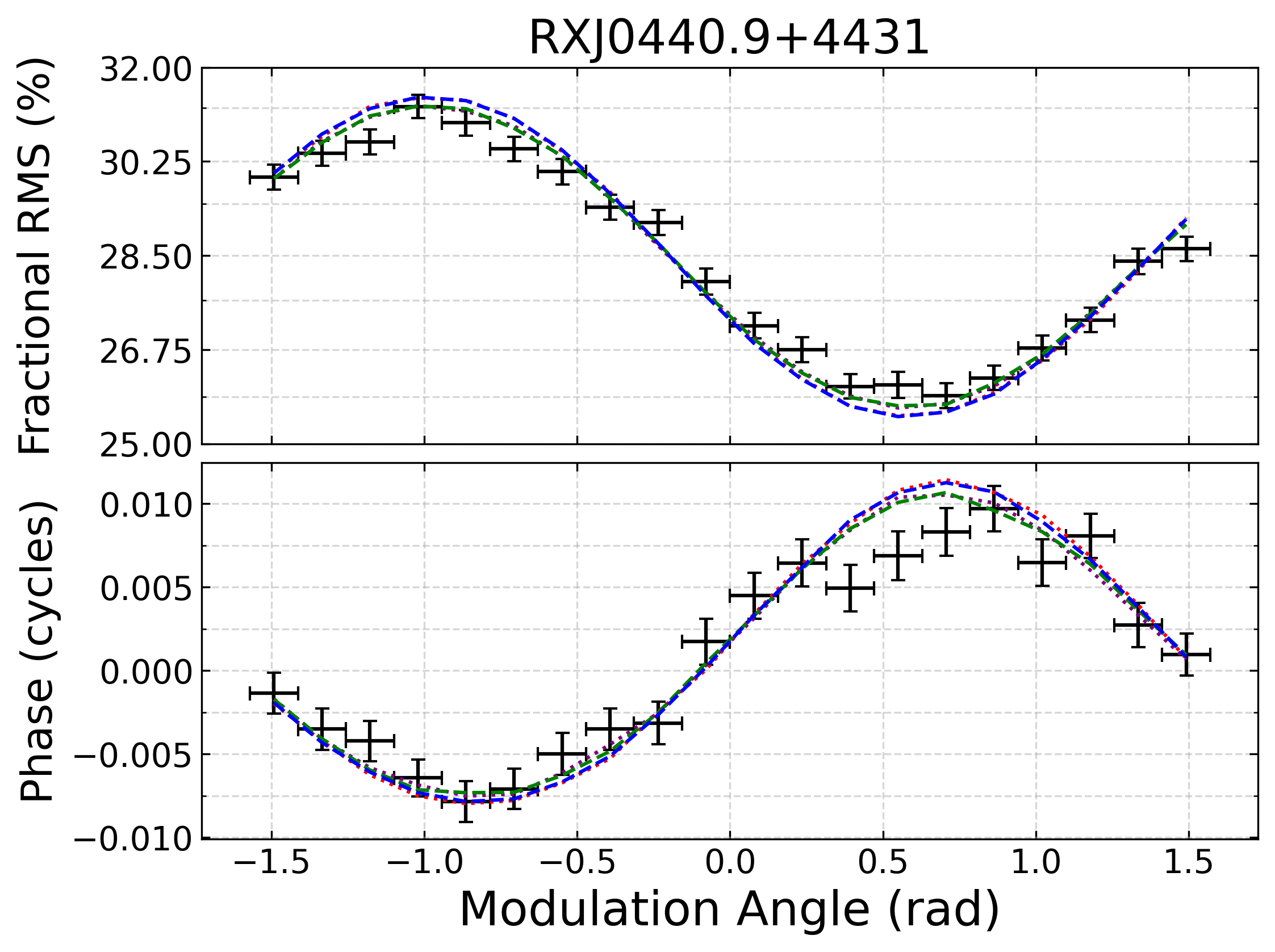}
        
        \label{fig:image1}
    \end{minipage}%
    \hfill
    \begin{minipage}{0.49\textwidth}
        \centering
        \includegraphics[width=1\linewidth]{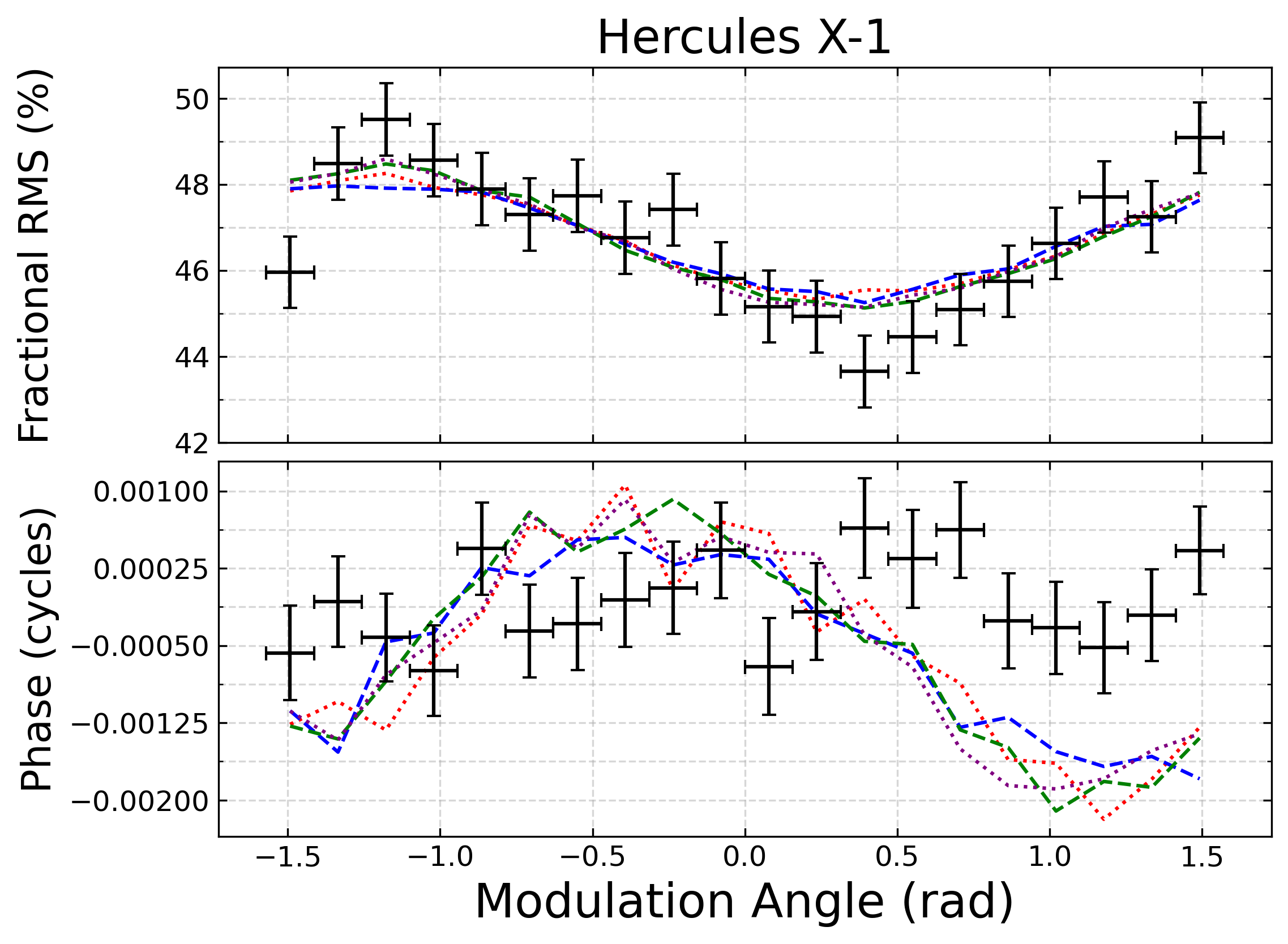}
        
        \label{fig:image2}
    \end{minipage}

        \caption{Simulated observations of \rxj (left) and \herx (right). The black data points represent the observed data. The red and purple dotted lines represent simulated data with signal injection that does not consider spurious polarisation and has a variable  and constant modulation factor respectively. The constant modulation  factor is of that used in the analytic calculation (see Section \ref{sec:analytic}).The blue and green dashed lines consider spurious polarisation with variable and constant modulation factor respectively. }
        
    \label{fig:sim data}
\end{figure*}
\\
Figure \ref{fig:sim data} compares the observed data to the four simulated scenarios.
The standard error on the mean between realisations is negligible and so is omitted from the plot. The red dotted lines do not consider spurious polarisation and have a variable modulation factor. The purple dotted lines do not consider spurious polarisation and uses a constant modulation factor. The blue dashed lines consider spurious polarisation and has a variable modulation factor. The green dashed lines consider spurious polarisation and have a constant modulation factor.

For \rxj, we see that each simulation set is in very strong agreement with each other and with the observed data. This shows that the effects of spurious polarisation have a negligible effect on the overall fractional rms and phase as a function of modulation angle, and similarly with the effect of a variable modulation factor on the fractional rms. For \herx we see that, for the fractional rms, the simulation sets again match the data. However, there is a small discrepancy between the simulation sets and the observed data for the lags, despite the lower signal-to-noise of this observation. We believe that this results either from the implicit assumption in our simulation that the pulse properties (I, Q and U as a function of pulse phase) are independent of energy, or from inaccuracies in our polynomial fit to the pulse properties. All other affects are accounted for in the simulation.

\section{Conclusions}
\label{sec:Conclusions}
The launch of IXPE has allowed us to test our Fourier-based method of fast stochastic X-ray polarimetry timing \citep{10.1093/mnras/stx1881, https://doi.org/10.48550/arxiv.2206.11671} for the \textit{first time} on real polarisation data. Here we have implemented the method by addressing several technical challenges. We have written an extension to the IXPE pipeline to enable the modulation angle of each event to be recovered, we have developed a framework to circumvent the effects of deadtime and constant spurious polarisation, and we have run Monte Carlo simulations to demonstrate that effects caused by rapid variability of the source spectral shape (namely variable spurious polarisation and modulation factor) are small.

We have tested our method on X-ray pulsars, which are the only sources with a `ground truth' of polarisation variability, provided by traditional phase-folding techniques. We show through a simple null hypothesis test that we detect polarisation variability over the pulse period in \rxj and Hercules X-1. In each case, a sinusoidal dependence of fractional rms and phase as a function of modulation angle is preferred statistically over no dependence, indicating the presence of polarisation variability. Moreover, we are able to reproduce the sinusoidal rms and phase modulations with a simple analytical model that can be used to measure the amplitude of PD and PA modulations. Our method is unique in that it is completely model-independent. At no point do we assume the phase of any one event, and so extension of our method to quasi-periodic and aperiodic variability is trivial.
Furthermore we can use our method on arbitrarily short timescales.

These properties give us the freedom to use this method on any fast stochastic X-ray polarisation variability, opening a new avenue into the study of XRBs and accretion processes. Perhaps the most obvious application is in the study of QPOs \citep{https://doi.org/10.48550/arxiv.2206.11671}. We now have the power to test for polarisation variability over the QPO frequency, which would provide smoking gun evidence for the LT precession model as the driver of the LF QPO \citep{Ingram_2019}.
The current best dataset for this test comes from IXPE observations of the recently discovered BH XRB Swift J1727.8-1613, which exhibits strong Type-C QPOs \citep{veledina2023,ingram2024trackingxraypolarizationblack}. We will therefore present an analysis of this dataset in future work (Ingram et al in prep). A search for a polarisation QPO in this dataset has already been conducted by \cite{Zhao_2024} using a phase-folding technique (the Hilbert-Huang transform). They report a negative detection, with upper limits on PD and PA QPO amplitude that we will explore in the context of the LT precession model in future work (Ewing et al in prep). Although the Hilbert-Huang transform has proven effective for phase-folding QPOs, we argue that our method has the advantage of being more agnostic of assumptions. It is also important to compare different methods. Indeed, comparison of the spectral-timing equivelent of these two methods has yielded interesting insights \citep{shui2024}, with the phase-folding method washing out variability in the second QPO harmonic compared with the Fourier method due to the stochastically variable phase difference between QPO harmonics \citep{Ingram2015}.

Furthermore, our method can be used to search for aperiodic variability in polarisation properties on arbitrarily short timescales. Such variability may be caused by, for example, propagation of accretion rate fluctuations from a less polarised region such as the corona to a more polarised region such as the jet \citep{Derosa_2019,https://doi.org/10.48550/arxiv.2206.11671}. In such a situation, a propagation lag could be detected even if the spectra from the two regions are identical, as long as the polarisation properties were different.
The field of X-ray reverberation mapping also serves to benefit from our method by measuring the lags between the highly polarised reflected rays and the weakly polarised corona \citep{https://doi.org/10.48550/arxiv.2206.11671}. Each of these lag types are expected to be small, but future missions such as the extended X-ray Timing and Polarimeter mission (eXTP) \citep{2019SCPMA..6229502Z} could provide the throughput required to carry out this analysis.


In conclusion, we have successfully implemented and tested the Fourier polarimetry timing method of \cite{ 10.1093/mnras/stx1881} for use with IXPE. Our method's applicability to any kind of polarisation variability, even over arbitrarily short timescales, opens a new discovery space, including potentially novel areas that have thus far not been predicted theoretically.

\section*{Acknowledgements}

M.E. and A.I. acknowledge support from the Royal Society. 
We thank Dr Victor Doroshenko for their contributions to data analysis.

\section*{Data Availability}
IXPE data are publicly available from HEASARC data archive (\url{https://heasarc.gsfc.nasa.gov)}.
 



\bibliographystyle{mnras}
\bibliography{bib} 




\appendix

\section{Level 1.5 event file pipeline}

In this Appendix, we describe our code to extract modulation angles in the sky frame, and write them to Level 1.5 event files.

\subsection{The IXPE pipeline}
\label{sec:ixpeSP}

In order to adequately describe how our code works, we must first introduce the details of the IXPE pipeline that produces Level 2 event files from Level 1 event files and attitude/callibration data. IXPE initially measures the modulation angle of each detected photon in the detector frame (i.e. the modulation angle is measured with respect to some fixed axis in the detector). Let us call this angle $\phi^{\rm det}_{\rm raw}$.  The superscript `det' denotes that the angle is in the detector reference frame, and the subscript `raw' denotes that the event has not yet been corrected for spurious polarisation. It is these raw modulation angles that are stored in the Level 1 event files (in the column entitled \texttt{DETPHI2}). We can define Stokes parameters for each event
\begin{equation}
    q_{\rm raw}^{\rm det} = 2\cos(2\phi_{\rm raw}^{\rm det}),~~~~~~~~u_{\rm raw}^{\rm det} = 2 \sin(2\phi_{\rm raw}^{\rm det}).
    \label{eqn:stokesrawdet}
\end{equation}


The IXPE pipeline first applies a correction for spurious polarisation
\begin{equation}
    q_{\rm cor}^{\rm det} =  q_{\rm raw}^{\rm det} - q_{\rm sp}^{\rm det},~~~~~~u_{\rm cor}^{\rm det} = u_{\rm raw}^{\rm det}- u_{\rm sp}^{\rm det},
\end{equation}
where $q_{\rm sp}^{\rm det}$ and $u_{\rm sp}^{\rm det}$ are interpolated from spurious polarisation maps for each DU that were constructed during the pre-launch calibration process (see Section \ref{sec:procedure} for more detail on the spurious polarisation maps).

It then applies a rotation by the detector roll angle, $\Delta$. The time-dependent roll angle is determined from the satellite attitude data (i.e. from the time-dependent orientation of the satellite). The 3 DUs are mounted onto the satellite in a circular configuration, such that contiguous DUs are separated by an angle of $120^\circ$. Thus for a given event, the difference between the $\Delta$ measured by DU N and the $\Delta$ measured by DU N-1 is exactly $120^\circ$. The sky frame event-by-event Stokes parameters are calculated from the following rotation
\begin{eqnarray}
    q^{\rm sky}_{\rm corr} &=& q^{\rm det}_{\rm cor} \cos( 2 \Delta) + u^{\rm det}_{\rm cor} \sin(2\Delta) \nonumber \\
    u^{\rm sky}_{\rm corr} &=& q^{\rm det}_{\rm cor} \sin(2\Delta) - u^{\rm det}_{\rm cor} \cos(2\Delta).
    \label{eqn:rot}
\end{eqnarray}
Note that the sign convention in the above results from the sky frame being mirrored with respect to the detector frame (similar to east and west being reversed when we look at the Earth or at the sky), and the factors of two arise because the roll angle is defined as a true angle periodic on a $360^\circ$ interval, whereas the modulation angles are associated with polarisation and thus are only periodic on a $180^\circ$ interval. It is the above sky frame Stokes parameters that are stored in the Level 2 event files (in the columns entitled \texttt{Q} and \texttt{U}).

\subsection{Method}
\label{sec:Deltamethod}

For our analysis, we need the raw modulation angle in the sky frame, $\phi_{\rm raw}^{\rm sky}$ for each event. This is not stored in the event files, but we can recover it by rotating the raw modulation angle in the detector frame $\phi_{\rm raw}^{\rm det}$, which is stored in the Level 1 event files. Following Equation \ref{eqn:rot}, the rotation is
\begin{equation}
    \phi^{\rm sky}_{\rm raw} = \Delta - \phi^{\rm det}_{\rm raw}.
    \label{eqn:phiskyrot}
\end{equation}
The roll angles are not stored in the event files, but we recover them from the following formula
\begin{equation}
    \Delta = \frac{1}{2} \arctan\left( \frac{ u^{\rm det}_{\rm raw} - u^{\rm det}_{\rm sp} } { q^{\rm det}_{\rm raw} - q^{\rm det}_{\rm sp} } \right) + \frac{1}{2} \arctan\left( \frac{ u^{\rm sky}_{\rm cor} } { q^{\rm sky}_{\rm cor} } \right).
    \label{eqn:Deltatan}
\end{equation}
We read $q^{\rm sky}_{\rm cor}$ and $u^{\rm sky}_{\rm cor}$ from the Level 2 event files, and calculate $q^{\rm det}_{\rm raw}$ and $u^{\rm det}_{\rm raw}$ from $\phi_{\rm raw}^{\rm det}$ using Equation \ref{eqn:stokesrawdet}. The spurious polarisation Stokes parameters used for each event, $q_{\rm sp}^{\rm det}$ and $u_{\rm sp}^{\rm det}$, are not stored in the event files, and thus we interpolate them from the spurious polarisation maps in the calibration database.

\subsection{Procedure}
\label{sec:procedure}

To implement the method described in the previous subsection, we read in the Level 1 and Level 2 event files for each of the three DUs. We match each Level 2 event to its corresponding Level 1 event by comparing time of arrival (note that some Level 1 events are filtered out before Level 2, but every Level 2 event has a corresponding Level 1 event). We then read the raw $x$ and $y$ detector position of each event from the Level 1 file (\texttt{ABSX} and \texttt{ABSY}) and the PI channel from the Level 2 file (\texttt{PI}).

To determine the spurious polarisation Stokes parameters for each event, we read in the 300 by 300 pixel spurious polarisation maps from the calibration database \citep{Rankin2022}. Each pixel in each map is labelled by x and y pixel numbers, $i_x$ and $i_y$, that both run from $1$ to $300$. There are six of these maps for each DU, with each one corresponding to a different energy channel: PI channels 51, 57, 66, 77, 92 and 148. For each event, we first select the correct pixel of the map to use by converting the $x$ and $y$ positions provided in millimeters in the Level 1 files to x and y pixel numbers using a simple linear conversion: $(x/{\rm mm}) = i_x \times 15 / 300 - 7.5$, $(y/{\rm mm}) = i_y \times 15 / 300 - 7.5$. Some events have $|x|$ or $|y|$ greater than $7.5$ mm, which is outside of the physical detector area. This is possible because the event position is reconstructed from the electron track and so has an uncertainty associated with it. We discard these events (they will eventually be discarded in any case because they are far outside of any reasonable source region).

We then interpolate on PI channel, using the PI channel read from the Level 2 event file. For events in PI channels 51-148, we linearly interpolate spurious polarisation Stokes parameters between channels. For events in channels below 51, we extrapolate and for channels above 148, we simply use the spurious polarisation Stokes parameters for channel 148. This exactly follows the procedure in the IXPE pipeline.

Once we have the spurious polarisation Stokes parameters for each event, $q_{\rm sp}^{\rm det}$ and $q_{\rm sp}^{\rm det}$, we combine them with the Level 2 Stokes parameters, $q_{\rm cor}^{\rm sky}$ and $q_{\rm cor}^{\rm sky}$, and the Level 1 modulation angle, $\phi_{\rm raw}^{\rm det}$, to obtain $\Delta$ from Equation \ref{eqn:Deltatan}. We then calculate the modulation angle in the sky frame, $\phi_{\rm raw}^{\rm sky}$, using Equation \ref{eqn:phiskyrot}.

To produce a Level 1.5 event file, we start from the Level 2 event file and append five new columns. The first, entitled \texttt{PHI}, contains $\phi_{\rm raw}^{\rm sky}$. The second has the title \texttt{QUAL}, which is a quality flag (described in the following subsection). The third new column has the title \texttt{DELTA}, and contains half the roll angle $\Delta/2$. The final two new columns, entitled \texttt{QSP} and \texttt{USP}, list the spurious polarisation Stokes parameters for each event.

\subsection{Tests}
\label{sec:tests}

We employ two diagnostic tests to check that our recovered spurious polarisation Stokes parameters and roll angles are the same as those calculated (but not subsequently stored) within the IXPE pipeline. First, we calculate
\begin{equation}
    \epsilon_2 = \sqrt{ (q_{\rm cor}^{\rm sky})^2 + (u_{\rm cor}^{\rm sky})^2 },
\end{equation}
and
\begin{equation}
    \epsilon_{\rm cor} = \sqrt{ (q_{\rm raw}^{\rm det}-q_{\rm sp}^{\rm det})^2 + (u_{\rm raw}^{\rm det}-u_{\rm sp}^{\rm det})^2 },
\end{equation}
for each event. Here, $\epsilon_2$ is calculated only from the Level 2 Stokes parameters, whereas $\epsilon_{\rm cor}$ is calculated from the Level 1 Stokes parameters and our interpolated spurious polarisation Stokes parameters. If our pipeline has produced exactly the same result as the IXPE pipeline, then we should find that $\epsilon_2 = \epsilon_{\rm cor}$ for every event, since the corrected Stokes parameters in the detector and sky frame are related by only a rotation.

We perform this test on the data from the May 2022 observation of Cygnus X-1. For this observation, there are 5,713,000 events in the DU1 Level 1.5 event file with well defined pixel numbers (i.e. we discard all events with $|x|>7.5$ mm or $|y|> 7.5$ mm). Of these, only 205 have $|\epsilon_2 - \epsilon_{\rm cor}| \geq 10^{-4}$. We find similar results for the other DUs, with a fraction $< 3.6 \times 10^{-5}$ of all events failing the test. We therefore simply discard the few events that fail the test. To do this, we define a quality flag that is unity if
$|\epsilon_2 - \epsilon_{\rm cor}|$ is smaller than some user defined tolerance, and zero otherwise. We record this quality flag in the new column \texttt{QUAL} of the Level 1.5 event files. For this paper, we use a tolerance of $10^{-4}$.

\begin{figure*}
    \centering
    \includegraphics[width=\columnwidth,trim=0.0cm 1.0cm 1.0cm 1.0cm,clip=true]{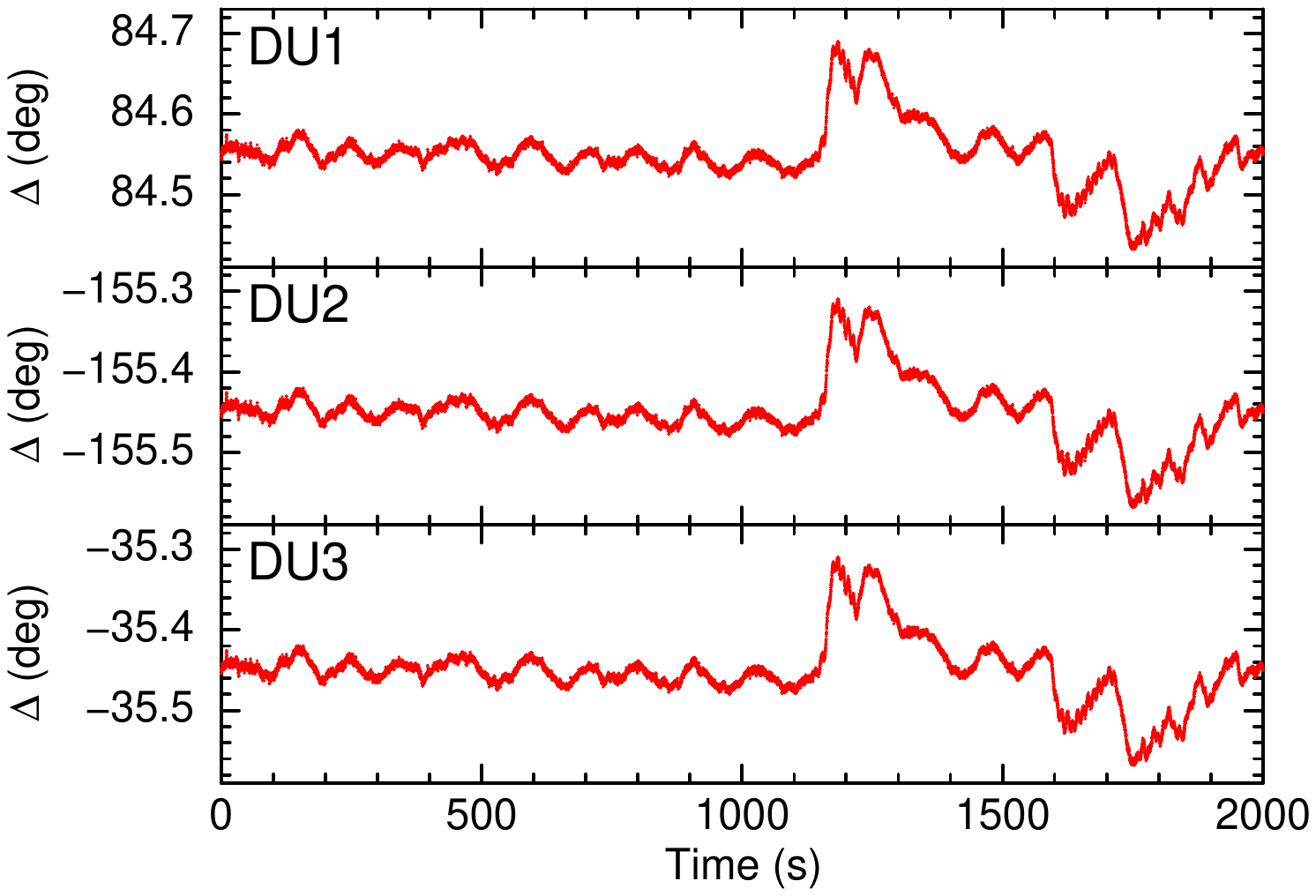}
    \caption{Roll angle derived by comparing each Level 2 event with the corresponding Level 1 event and subtracting the spurious polarisation interpolated for that event. Only events that pass the $\epsilon_{\rm cor} = \epsilon_2$ test are plotted.}
\label{fig:Delta}
\end{figure*}

The second test is a `sanity check' on the derived $\Delta$ values. We know that the different DUs are separated by $120^\circ$, and we also know that variations in $\Delta$ should be small ($\sim$arcminutes) and fairly gentle in time. Specifically, we know that there are three dithering periods: 107, 127 and 900 s, which we should be able to roughly see in a time series of $\Delta$. Fig \ref{fig:Delta} shows such a time series for the same observation of Cygnus X-1. We see that the $\Delta$ values for different DUs are indeed separated by $120^\circ$, the variations in $\Delta$ are small, and we can clearly see a $\sim 100$ s periodicity.

\section{Light curve correction for constant spurious polarisation}
\label{sec:ncorr}

Here we explicitly derive the global decoupling correction for spurious polarisation. From Equation (\ref{eqn:fullf}), we see that the number of counts in the $j^{\rm th}$ modulation angle bin (summed over all time bins) is
\begin{equation}
    N(\phi_j) = N~\frac{\Delta\phi}{\pi} \bigg[ 1 + ( q + q_{\rm sp}) \cos(2\phi_j) + ( u + u_{\rm sp}) \sin(2\phi_j) \bigg].
\end{equation}
We wish to isolate
\begin{equation}
    N_{\rm corr}(\phi_j) = N~\frac{\Delta\phi}{\pi} \bigg[ 1 + q \cos(2\phi_j) + u \sin(2\phi_j) \bigg],
\end{equation}
and thus the required correction is
\begin{equation}
    N_{\rm corr}(\phi_j) = N(\phi_j) - \frac{N}{J} \bigg[ 1 + q_{\rm sp} \cos(2\phi_j) + u_{\rm sp} \sin(2\phi_j) \bigg],
\end{equation}
where we have used $\Delta\phi = \pi/J$. Note that $q_{\rm sp}$ and $u_{\rm sp}$ are calculated from Equation (\ref{eqn:qsp}), where each set of event by event spurious polarisation Stokes parameters is selected from the characteristics of the corresponding real event (DU, chip position and energy channel; see Appendix \ref{sec:procedure} for more details). This is the global decoupling correction to be applied to the total histogram of counts versus modulation angle summed over all time bins.

We apply the same constant spurious polarisation correction to each time bin of our modulation angle selected light curves. In this case, $N(t_i,\phi_j)$ is the number of counts in the $i^{\rm th}$ time bin and $j^{\rm th}$ modulation angle bin, such that $N(\phi_j) = \sum_{i=1}^I N(t_i,\phi_j)$. Assuming that the correction is constant in time leads to Equation (\ref{eq:ncorr}), since the average counts per time bin is $N/I$ if there are $N$ counts summed over $I$ time bins.


\bsp	
\label{lastpage}
\end{document}